\documentclass[reprint,twocolumn,pra,aps,superscriptaddress]{revtex4-1}%
\usepackage{amssymb}
\usepackage{graphicx,float}
\usepackage{amsmath}
\usepackage{amsfonts}
\usepackage{accents}
\usepackage{color}

\DeclareMathOperator{\Tr}{Tr}
\newcommand{\ket}[1]{|#1 \rangle}
\newenvironment{psmallmatrix}
  {\left(\begin{smallmatrix}}
  {\end{smallmatrix}\right)}

\begin{document}
\title{Quantum Illumination with a generic Gaussian source}
\author{Athena Karsa}
\affiliation{Department of Computer Science, University of York, York YO10 5GH, UK}
\author{Gaetana Spedalieri}
\affiliation{Research Laboratory of Electronics, Massachusetts Institute of Technology,
Cambridge MA 02139, USA}
\affiliation{Department of Computer Science, University of York, York YO10 5GH, UK}
\author{Quntao Zhuang}
\affiliation{Department of Electrical and Computer Engineering, University of Arizona,
Tucson, Arizona 85721, USA}
\affiliation{James C. Wyant College of Optical Sciences, University of Arizona, Tucson, AZ
85721, USA}
\author{Stefano Pirandola}
\affiliation{Department of Computer Science, University of York, York YO10 5GH, UK}
\affiliation{Research Laboratory of Electronics, Massachusetts Institute of Technology,
Cambridge MA 02139, USA}

\begin{abstract}
With the aim to loosen the entanglement requirements of quantum illumination,
we study the performance of a family of Gaussian states at the transmitter,
combined with an optimal and joint quantum measurement at the receiver. We
find that maximal entanglement is not strictly necessary to achieve quantum
advantage over the classical benchmark of a coherent-state transmitter, in
both settings of symmetric and asymmetric hypothesis testing. While performing this quantum-classical comparison, we also investigate a suitable regime of parameters for potential short-range radar (or scanner) applications.

\end{abstract}
\maketitle

\section{Introduction}

Hypothesis testing (HT)~\cite{lehmann2006testing} and quantum hypothesis
testing (QHT)~\cite{Helstrom} play crucial roles in information~\cite{Cover}
and quantum information theory~\cite{watrous2018quantum}. HT has fundamental
links to both communication and estimation theory, ultimately underlying the
task of radar detection~\cite{richards2005fundamentals} which has been
extended to the quantum realm by the protocol of quantum illumination
(QI)~\cite{lloyd2008enhanced,tan2008quantum} and, more precisely, by the model
of microwave quantum illumination~\cite{barzanjeh2015microwave} (see
Ref.~\cite{reviewSENSING}\ for a recent review on these topics). The simplest
scenario of both HT and QHT is that of a binary decision, so that they are
reduced to the statistical discrimination between just two hypotheses (null,
$H_{0}$, and alternative, $H_{1}$).

At its most basic level a quantum radar is a task of binary QHT. The two
alternate hypotheses are encoded in two quantum channels in through which a
signal mode is sent. Depending on the presence or not of a target, the initial
state of signal mode undergoes different transformations which result in two
different quantum states at the output. Final detection is then reduced to
distinguishing between these two possible quantum states. The ability to do
this accurately, with a low probability of error, directly relates to an
ability to determine the correct result. This fundamental mechanism can then
be easily augmented with geometrical ranging arguments which account for the
quantification of the round-trip time from the target, i.e., its distance.

While QI radars may potentially achieve the best performances~\cite{FFSFG},
they require the generation of a large number of entangled states which may be
a demanding task, especially if we consider the microwave regime. At the same
time, the definition itself of quantum radar may be generalized beyond QI to
any model that exploits a quantum part or device to beat the performance of a
corresponding classical radar in the same conditions of energy, range etc.
Driven by these ideas, we progressively relax the entanglement requirements of
QI and we study the corresponding detection performances to the point where
the source becomes just-separable, i.e., a maximally-correlated separable
state. It is worth noting that, although Gaussian entanglement is the main resource of QI, previous literature has also considered the use of separable non-Gaussian sources with non-positive P-representations, finding an advantage over a restricted classical benchmark~\cite{lopaevaexp}. More generally, quantum correlations beyond entanglement have also been considered for a number of other quantum information and computation tasks~\cite{agudelo,sperling,shahandeh,spedalieri}. However, our current study is specifically focused on Gaussian states because they are, so far, the only sources showing a quantum advantage over the best classical benchmark. The analysis is done in the setting of symmetric and asymmetric QHT. In particular we show how a quantum advantage can still be achieved with less entangled sources, especially in a scenario of very short-range target detection.

\section{General quantum-correlated source\label{SECgen}}

Following Gaussian QI, we consider a source modelled as a two-mode Gaussian
state~\cite{RMP}, comprising a signal ($s$) mode, sent out to some target
region, and an idler ($i$) mode, retained at the source for later joint
measurement. Each of these modes has $N_{s}$ mean number of photons. However,
instead of using a two-mode squeezed vacuum (TMSV) state~\cite{RMP} as in QI,
we can employ a generic zero-mean Gaussian state whose covariance matrix (CM)
takes the following block form
\begin{align}
\mathbf{V}_{si}^{gen}  &  =%
\begin{pmatrix}
S & C\\
C & S
\end{pmatrix}
\oplus%
\begin{pmatrix}
S & -C\\
-C & S
\end{pmatrix}
,\label{eq1}\\
S  &  :=N_{s}+1/2,\label{eq2}\\
0  &  \leq C\leq\sqrt{S^{2}-1/4}=\sqrt{N_{s}(N_{s}+1)},\label{eq3}%
\end{align}
where the direct sum operator $\oplus$ acts on two matrices $A$ and $B$ such that $A \oplus B = \begin{psmallmatrix} A & 0 \\ 0 & B \end{psmallmatrix}$. The terms in the leading diagonal, $S:=N_{s} + 1/2$, quantify the amount of thermal noise within each of the local modes, $S$ and $I$, while the covariance, $C$, quantifying the correlations between these two modes, may take any value within the range given by Eq.~(\ref{eq3}). Mathematically, these are the second-order statistical moments of the quantum state (see Ref.~\cite{RMP} for more details).

At maximal quantum correlations we have $C=C_{q}:=\sqrt{N_{s}(N_{s}+1)}$,
corresponding to the TMSV state, while the case $C=C_{d}:=N_{s}$ renders the
state just-separable~\cite{EntBreak}. At this border point, the state is not
entangled but it still has quantum correlations~\cite{ModiDiscord}. In fact, its quantum discord
is maximal among the states within the range $C\leq C_{d}$ and is equal to its
Gaussian discord~\cite{OptDisc} (therefore computable using
Refs.~\cite{Gdis1,Gdis2}). This kind of source has already played a
non-trivial role in other problems of quantum information theory, e.g., as
candidate separable state in relative entropy bounds for the two-way quantum
capacities of bosonic Gaussian channels~\cite{PLOB}. In our work, we will then
relax the QI\ model by studying the performance of the source in
Eqs.~(\ref{eq1})-(\ref{eq3}), up to the border case of $C=C_{d}$.

Let us now look at the output state at the receiver. Two hypotheses exist for
the experiment's outcome:

\begin{description}
\item[$H_{0}:$] Target is absent, so that the return signal is a noisy
background modelled as a thermal state with mean number of photons per mode
$N_b\gg1$.

\item[$H_{1}:$] Target is present with reflectivity $\kappa\ll1$, so that a
proportion of signal modes is reflected back to the transmitter. In this
high-loss regime, the return signal is combined with a very strong background
with mean photons per mode $N_b/(1-\kappa)$.
\end{description}

\noindent The joint state of our returning ($r$) mode and the retained idler
is given by, under $H_{0}$ and $H_{1}$, respectively:
\begin{equation}
\mathbf{V}_{ri}^{(0)}=%
\begin{pmatrix}
B & 0\\
0 & S
\end{pmatrix}
\oplus%
\begin{pmatrix}
B & 0\\
0 & S
\end{pmatrix}
,
\label{RI0}
\end{equation}%
\begin{equation}
\mathbf{V}_{ri}^{(1)}=%
\begin{pmatrix}
A & \sqrt{\kappa}C\\
\sqrt{\kappa}C & S
\end{pmatrix}
\oplus%
\begin{pmatrix}
A & -\sqrt{\kappa}C\\
-\sqrt{\kappa}C & S
\end{pmatrix}
,
\label{RI1}
\end{equation}
where $B:=N_b+1/2$ and $A:=\kappa N_{s}+B$. For an arbitrary Gaussian state with leading diagonal entries $a$ and $b$, separability corresponds to the off-diagonal term $c \leq c_d :=\sqrt{(a-1/2)(b-1/2)}$~\cite{PLOB}. For each of these output quantum states, conditional on $H_0$ and $H_1$, we have that $0 \leq  \sqrt{N_b N_{s}}$ and $ \sqrt{\kappa} C \leq \sqrt{(\kappa N_{s} + N_b) N_{s}}$, for small $\kappa$, respectively, thus the separability criterion is always satisfied and neither of these states are entangled.

In the absence of the idler, the best strategy is to use coherent states. This
is a semi-classical design which is used as a classical benchmark in quantum
information to evaluate the effective performance of quantum-correlated
sources~\cite{tan2008quantum,reviewSENSING}. Let us work within the formalism of creation, $\hat{a}^{\dag}$, and annihilation, $\hat{a}$, operators for bosonic modes defined by
\begin{equation}
\hat{a}^{\dag} \ket{n}=\sqrt{n+1} \ket{n+1} ,
\end{equation}
\begin{equation}
\hat{a} \ket{n}=\sqrt{n} \ket{n-1} ,
\end{equation}
where $\ket{n}$ is a Fock state (an eigenstate of the photon-number operator $\hat{n} =\hat{a}^{\dag} \hat{a}$). Letting $\hat{a}_{s}$ be the
annihilation operator for the signal mode prepared in the coherent state
$\ket{\sqrt{N_{s}}}$ (satisfying the eigenvalue equation $\hat{a}_{s} \ket{\sqrt{N_{s}}}= \sqrt{N_{s}}  \ket{\sqrt{N_{s}}}$), we send such a mode to some target region. Under
$H_{0}$ the return signal, with annihilation operator $\hat{a}_r$, is equal to that of the background which is in a thermal state with mean photons per mode $N_b$, i.e.,  $\hat{a}_r=\hat{a}_b$. The state has mean vector of
zero and CM $(N_b+1/2)\mathbf{1}_{2}$, where $\mathbf{1}_{2}$ is the $2 \times 2$ identity matrix. Under $H_{1}$ the target is present and reflects a small proportion of our signal back. This is mixed with with the background radiation such that our return takes the
form $\hat{a}_r=\sqrt{\kappa}\hat{a}_{s}+\sqrt{1-\kappa}\hat{a}_b$, where
$\kappa\in(0,1)$ and the background has mean photons per mode $N_b%
/(1-\kappa)$. This corresponds to a displaced thermal state with mean vector
$(\sqrt{2\kappa N_{s}},0)$ and CM $(N_b+1/2)\mathbf{1}_{2}$.

\section{Hypothesis testing for quantum radar detection\label{secHT}}

Radar detection requires successful distinguishing between the two
alternatives $H_{0}$ and $H_{1}$, which happens with detection probability
$P_{\text{d}}:=P(H_{1}|H_{1})$. There are two types of error which may occur:
type-I (false alarm) error $P_{\text{fa}}=P(H_{1}|H_{0})$, where we
incorrectly reject the null hypothesis, and type-II (missed detection) error
$P_{\text{md}}(H_{0}|H_{1})$, where we incorrectly reject the alternative
hypothesis. The optimization of these probabilities can be carried out in a
range of ways based on the potentially situation-dependent rules one wishes to
follow for decision making. That is, one can associate with each error type a
cost. For example, considering the result of a diagnostic test then it is
clear that the risk associated with receiving a false negative (type-II) could
far outweigh that associated with a false positive (type-I). In such scenarios
one may consider \emph{asymmetric} testing in order to take in account these
discrepancies. On the other hand, a \emph{symmetric} approach may be used if
one's aim is to obtain a global minimization over all errors, irrespective of
their origin. In this case, one considers the minimization of the average
error probability
\begin{equation}
P_{\text{err}}:=P(H_{0})P(H_{1}|H_{0})+P(H_{1})P(H_{0}|H_{1}),\label{avgP}%
\end{equation}
where $P(H_{0})$ and $P(H_{1})$ are the prior probabilities associated with
the two hypotheses.

In the following subsections, we briefly review the main tools for symmetric
and asymmetric QHT. We will use these tools for the results of the next sections.

\subsection{Review of symmetric detection}\label{SECsymdet}

In symmetric QHT, the average error probability $P_{\text{err}}$ of
Eq.~(\ref{avgP}) is minimized. Consider $M$ identical copies $\hat{\rho}%
_{i}^{\otimes M}$ of the state $\hat{\rho}_{i}$ encoding the classical
information bit $i\in\{0,1\}$. The optimal measurement for the discrimination
is the dichotomic positive-operator valued measure
(POVM)~\cite{helstrom1969quantum} $E_{0}=\Pi(\gamma_{+})$, $E_{1}=1-\Pi
(\gamma_{+})$, where $\Pi(\gamma_{+})$ is the projector on the positive part
$\gamma_{+}$ of the non-positive Helstrom matrix $\gamma:=\hat{\rho}_{0}^{\otimes M}-\hat{\rho}_{1}^{\otimes M}$. This allows for $\hat{\rho}_{0}$ and $\hat{\rho}_{1}$ to be
discriminated with a \emph{minimum} error probability given by the Helstrom
bound, $P_{\text{err}}^{\mathrm{min}}=\left[  1-D(\hat{\rho}_{0}^{\otimes M},\hat{\rho
}_{1}^{\otimes M})\right]  /2$, where $D$ is the trace distance~\cite{watrous2018quantum}.


Because this is difficult to compute analytically, the Helstrom bound is often
replaced with approximations such as the quantum Chernoff bound (QCB)~\cite{QCB},
\begin{equation}
P_{\text{err}}^{\mathrm{min}}\leq P_{\text{err}}^{\mathrm{QCB}}:=\frac{1}%
{2}\left(  \inf_{0\leq s\leq1}C_{s}\right)^{M}  ,~~C_{s}:=\Tr\left(  \hat{\rho
}_{0}^{s}\hat{\rho}_{1}^{1-s}\right).\label{QCBtext}
\end{equation}
Minimization of the $s$-overlap $C_{s}$ occurs over all $0\leq s\leq1$.
Forgoing minimization and setting $s=1/2$ one defines a simpler, though
weaker, upper bound, also known as the quantum Bhattacharyya bound (QBB)~\cite{RMP}%
\begin{equation}
P_{\text{err}}^{\mathrm{QBB}}:=\frac{1}{2}\Tr\left(  \sqrt{\hat{\rho}_{0}%
}\sqrt{\hat{\rho}_{1}}\right)^{M}.\label{QBBtext}
\end{equation}
In the case of Gaussian states, we can compute these quantities by means of
closed analytical formulas~\cite{pirandola2008computable}.

Consider $N$ bosonic modes with quadratures $\mathbf{\hat{x}}=\left(  \hat
{q}_{1},\dots,\hat{q}_{N},\hat{p}_{1},\dots,\hat{p}_{N}\right)  ^{T}$ and
associated symplectic form
\begin{equation}
\mathbf{\Omega}=%
\begin{pmatrix}
0 & 1\\
-1 & 0
\end{pmatrix}
\otimes\mathbf{1}_{N},
\end{equation}
where $\mathbf{1}_{N}$ is the $N\times N$ identity matrix. Then consider two
arbitrary $N$-mode Gaussian states, $\hat{\rho}_{0}(\mathbf{x}_{0}%
,\mathbf{V}_{0})$ and $\hat{\rho}_{1}(\mathbf{x}_{1},\mathbf{V}_{1})$, with
mean $\mathbf{x}_{i}$ and CM $\mathbf{V}_{i}$. We can write the following
Gaussian formula for the $s$-overlap of the quantum Chernoff bound~\cite{pirandola2008computable}%
\begin{equation}
C_{s}=2^{N}\sqrt{\frac{\det\mathbf{\Pi}_{s}}{\det\mathbf{\Sigma}_{s}}}%
\exp\left(  -\frac{\mathbf{d}^{T}\mathbf{\Sigma}_{s}^{-1}\mathbf{d}}%
{2}\right)  ,
\end{equation}
where $\mathbf{d}=\mathbf{x}_{0}-\mathbf{x}_{1}$. Here $\mathbf{\Pi}_{s}$ and
$\mathbf{\Sigma}_{s}$ are defined as
\begin{equation}
\mathbf{\Pi}_{s}:=G_{s}(\mathbf{V}_{0}^{\oplus})G_{1-s}(\mathbf{V}_{1}%
^{\oplus}),
\end{equation}
\vspace{-0.5cm}
\begin{equation}
\mathbf{\Sigma}_{s}:=\mathbf{S}_{0}\left[  \Lambda_{s}\left(  \mathbf{V}%
_{0}^{\oplus}\right)  \right]  \mathbf{S}_{0}^{T}+\mathbf{S}_{1}\left[
\Lambda_{1-s}\left(  \mathbf{V}_{1}^{\oplus}\right)  \right]  \mathbf{S}%
_{1}^{T},
\end{equation}
introducing the two real functions
\begin{equation}
G_{s}(x)=\frac{1}{(x+1/2)^{s}-(x-1/2)^{s}},
\end{equation}
\vspace{-0.5cm}
\begin{equation}
\Lambda_{s}(x)=\frac
{(x+1/2)^{s}+(x-1/2)^{s}}{(x+1/2)^{s}-(x-1/2)^{s}},
\end{equation}
calculated over the Williamson forms $\mathbf{V}_{i}^{\oplus}%
:=\mathbf{\bigoplus}_{k=1}^{N}\nu_{i}^{k}\mathbf{1}_{2}$, where $\mathbf{V}%
_{i}^{\oplus}\mathbf{=S}_{i}\mathbf{\mathbf{V}}_{i}^{\oplus}\mathbf{S}_{i}%
^{T}$ for symplectic $\mathbf{S}_{i}$\ and $\nu_{i}^{k}\geq1/2$ are the
symplectic spectra~\cite{serafini2003symplectic,pirandola2009correlation}.

\subsection{Review of asymmetric detection}

In asymmetric QHT, we wish to minimize one type of error as much as possible
while allowing for some flexibility on the other. Consider again $M$ identical
copies of the state $\hat{\rho}_{i}$ ($\hat{\rho}_{i}^{\otimes M}$), encoding
the classical bit $i\in\{0,1\}$. As in the symmetric case, the optimal choice
of measurement is a dichotomic POVM $\{E_{0},E_{1}\}$. From the binary
outcome, we can define the two types of error, i.e., the type-I (false alarm)
error
\begin{equation}
P_{\text{fa}}:=P(H_{1}|H_{0})=\Tr\left(  E_{1}\hat{\rho}_{0}^{\otimes
M}\right)  ,
\end{equation}
and the type-II (missed detection) error
\begin{equation}
P_{\text{md}}:=P(H_{0}|H_{1})=\Tr\left(  E_{0}\hat{\rho}_{1}^{\otimes
M}\right)  .
\end{equation}
These probabilities are dependent on the number $M$ of copies and, for $M\gg1$,
they both tend to zero, i.e.,
\begin{equation}
P_{\text{fa}}\simeq e^{-\alpha_{R}M},~P_{\text{md}}\simeq e^{-\beta_{R}M},
\end{equation}
where we define the `error-exponents' or `rate limits' as
\begin{equation}
\alpha_{R}=-\lim_{M\rightarrow+\infty}\frac{1}{M}\ln P_{\text{fa}},
\end{equation}
\vspace{-0.5cm}
\begin{equation}
\beta_{R}=-\lim_{M\rightarrow+\infty}\frac{1}{M}\ln P_{\text{md}}.
\end{equation}

It is not possible to make both error probabilities arbitrarily small
simultaneously. Instead we place a relatively loose constraint $P_{\text{fa}%
}<\epsilon$ on the type-I error, allowing us more freedom to minimize
$P_{\text{md}}$. The quantum Stein's
lemma~\cite{hiai1991proper,ogawa2005strong} tells us that the quantum relative
entropy $D\left(  \hat{\rho}_{0}||\hat{\rho}_{1}\right)
=\Tr [\hat{\rho}_0(\ln \hat{\rho}_0 - \ln \hat{\rho}_1)]$ between two quantum
states, $\hat{\rho}_{0}$ and $\hat{\rho}_{1}$, is the optimal decay rate for
the type-II error probability, given some fixed constraint on the type-I error
probability. Further, if the type-II error tends to 0 with an exponent larger
than $D\left(  \hat{\rho}_{0}||\hat{\rho}_{1}\right)  $, then the type-I error
converges to 1~\cite{ogawa2005strong}. (Note that an alternative approach
based on the quantum Hoeffding bound~\cite{QHBound} is not considered here,
but it could be explored using the Gaussian formulas developed in
Ref.~\cite{GaeGAUSS}).

Refinement of quantum Stein's lemma has been provided by considering the
second order (in $M$) asymptotics~\cite{li2014second} to account for the
discontinuity observed in the type-I error probability, jumping sharply from 0
to 1, when the type-II error probability increases past the value set by
$D\left(  \hat{\rho}_{0}||\hat{\rho}_{1}\right)  $. Tracking the type-II error
exponent to second order depth, that is to order $\sqrt{M}$, allows one to
define the quantum relative entropy variance
\begin{equation}
V\left(  \hat{\rho}_{0}||\hat{\rho}_{1}\right)
=\Tr [\hat{\rho}_0(\ln \hat{\rho}_0 - \ln \hat{\rho}_1)^2]-[D\left(  \hat
{\rho}_{0}||\hat{\rho}_{1}\right)  ]^{2},
\end{equation}
and in turn establish that the optimal type-II (missed detection) error
probability, for sample size $M$, takes the exponential
form~\cite{li2014second}
\begin{equation}
\begin{split}
P_{\text{md}}=\exp\Big\{  -\Big[  MD\left(  \hat{\rho}_{0}||\hat{\rho}%
_{1}\right) &+\sqrt{MV\left(  \hat{\rho}_{0}||\hat{\rho}_{1}\right)  }%
\Phi^{-1}(\epsilon) \\& + \mathcal{O}(\log M)\Big]  \Big\}
,\label{typeIIerrorexpform}%
\end{split}
\end{equation}
where $\epsilon\in(0,1)$ bounds $P_{\text{fa}}$\ and
\begin{equation}
\Phi(y):=\frac{1}{\sqrt{2\pi}}\int_{-\infty}^{y}dx\exp\left(  -x^{2}/2\right)
\end{equation}
is the cumulative of a normal distribution. More precisely, for finite
third-order moment (as in the present case) and sufficiently large $M$, we may
write the upper bound~\cite[Theorem~5]{li2014second}%
\begin{equation}
\begin{split}
P_{\text{md}}\leq \tilde{P}_{\text{md}}:=& \exp\Big\{  -\Big[  MD\left(
\hat{\rho}_{0}||\hat{\rho}_{1}\right)  \\&+\sqrt{MV\left(  \hat{\rho}_{0}%
||\hat{\rho}_{1}\right)  }\Phi^{-1}(\epsilon)+\mathcal{O}(1)\Big]  \Big\}
.\label{ROCub}%
\end{split}
\end{equation}

We can write explicit formulas for the relative entropy $D\left(  \hat{\rho
}_{0}||\hat{\rho}_{1}\right)  $\ and the relative entropy variance $V\left(
\hat{\rho}_{0}||\hat{\rho}_{1}\right)  $ of two arbitrary $N$-mode Gaussian
states, $\hat{\rho}_{0}(\mathbf{x}_{0},\mathbf{V}_{0})$ and $\hat{\rho}%
_{1}(\mathbf{x}_{1},\mathbf{V}_{1})$. The first one is given by~\cite{PLOB}
\begin{equation}
D\left(  \hat{\rho}_{0}||\hat{\rho}_{1}\right)  =-\Sigma\left(  \mathbf{V}%
_{0},\mathbf{V}_{0}\right)  +\Sigma\left(  \mathbf{V}_{0},\mathbf{V}%
_{1}\right)  ,
\end{equation}
where we have defined the function
\begin{equation}
\Sigma\left(  \mathbf{V}_{0},\mathbf{V}_{1}\right)  =\frac{\ln\mathrm{det}%
\left(  \mathbf{V}_{1}+\frac{i\mathbf{\Omega}}{2}\right)  +\Tr\left(
\mathbf{V}_{0}\mathbf{G}_{1}\right)  +\delta^{T}\mathbf{G}_{1}\delta}{2},
\end{equation}
with $\delta=\mathbf{x}_{0}-\mathbf{x}_{1}$ and $\mathbf{G}_{1}%
=2i\boldsymbol{\Omega}\coth^{-1}\left(  2i\mathbf{V}_{1}\boldsymbol{\Omega
}\right)  $ being the Gibbs matrix~\cite{BanchiPRL}. The second one is given
by%
\begin{equation}
V\left(  \hat{\rho}_{0}||\hat{\rho}_{1}\right)  =\frac{\Tr\left[
(\mathbf{\Gamma}\mathbf{V}_{0})^{2}\right]  }{2}+\frac{\Tr\left[
(\mathbf{\Gamma}\mathbf{\Omega})^{2}\right]  }{8}+\delta^{T}\mathbf{G}%
_{1}\mathbf{V}_{0}\mathbf{G}_{1}\delta,
\end{equation}
where $\mathbf{\Gamma}=\mathbf{G}_{0}-\mathbf{G}_{1}$~\cite{RevQKD} (see also Ref.~\cite{LaurenzaBounds}).

\section{Quantum radar detection with general source}

Using the generic quantum-correlated Gaussian source of Sec.~\ref{SECgen}\ and the
tools for symmetric and asymmetric QHT of Sec.~\ref{secHT}, we study the
performance of a relaxed QI protocol, clarifying how much entanglement is
needed to beat the semi-classical benchmark of the coherent-state transmitter
under symmetric testing. Then, in the setting of asymmetric testing, we repeat
the study in terms of the receiver operating characteristic (ROC), where the
mis-detection probability is plotted versus the false alarm probability.

\subsection{Symmetric detection with general source}\label{gensource}

Let us assume the typical conditions of QI, which are low-reflectivity
$\kappa\ll1$, high thermal-noise $N_b\gg1$, low photon number per mode
$N_{s}\ll1$. It is then known that, using a TMSV state, the minimum error
probability satisfies~\cite{tan2008quantum}
\begin{equation}
P_{\text{err}}^{\text{TMSV}}\leq e^{-M\kappa N_{s}/N_b}%
/2.\label{bhattaqibound}%
\end{equation}
This is computed using the quantum Bhattacharyya bound, it is exponentially
tight in the limit of large $M$, and it is also known to be achieved by the
sum-frequency-generation receiver of Ref.~\cite{FFSFG}. Its error-rate
exponent has a factor of 4 advantage over the same bound computed over a
coherent-state transmitter in the same conditions, for which we
have~\cite{tan2008quantum}%
\begin{equation}
P_{\text{err}}^{\text{CS}}\leq e^{-M\kappa N_{s}/4N_b}%
/2.\label{coherentchernoffbound2}%
\end{equation}

In order to extend Eq.~(\ref{bhattaqibound}) to the error probability for a
generic Gaussian source, let us start with single probing $M=1$, assuming the usual
limits $\kappa\ll1$, $N_b\gg1$ and $N_{s}\ll1$. The quantum Bhattacharyya bound
takes the form
\begin{equation}
P_{\text{err}}^{\text{gen}}\leq e^{-\kappa N_{s}g_{C}(N_{s})/N_b}/2,
\label{regenbound}
\end{equation}
where the function $g_{C}(N_{s})$ is proportional to $C^{2}$ (see Appendix \ref{App1} for details), i.e., the amount
of correlations existing between the signal and idler modes. Demanding the
equivalence of exponents in the TMSV\ limit $C\rightarrow C_{q}$, we find that
the quantum Bhattacharrya bound for $M$ probings becomes
\begin{equation}
P_{\text{err}}^{\text{gen}}\leq e^{-M\kappa N_{s}C^{2}/N_bC_{q}^{2}}/2.
\label{gensourcebound}%
\end{equation}

By comparing Eqs.~(\ref{gensourcebound}) and~(\ref{coherentchernoffbound2}),
we see that a quantum-correlated transmitter beats the coherent state
transmitter if $P_{\text{err}}^{\text{gen}}\leq P_{\text{err}}^{\text{CS}}$
which means%
\begin{equation}
\frac{C^{2}}{C_{q}^{2}}\geq\frac{1}{4}\Rightarrow C\geq\frac{1}{2}\sqrt
{N_{s}(N_{s}+1)}. \label{beatclassrequirement}%
\end{equation}
Thus, according to the quantum Bhattacharyya bound, the quadrature correlations required to outperform the semi-classical
benchmark is half the value of those of a TMSV state. At the separable limit
$C=N_{s}$ the relation is only satisfied for $N_{s}\geq1/3$ which contradicts
the assumption $N_{s}\ll1$ (a similar analysis holds if we relax the assumption
of $N_{s}\ll1$). Therefore, according to the quantum Bhattacharyya bound, the employment of a source at the separable limit is not
capable of beating coherent states under symmetric testing.

\subsection{Asymmetric detection with general source}

Let us compute the quantum relative entropy and the quantum relative entropy
variance for the quantum-correlated transmitter of Eqs.~(\ref{eq1}%
)-(\ref{eq3}). Though the full expressions for these quantities are far too
long to display here, we evaluate them to first order in $N_b$ by taking an
asymptotic expansion for large $N_b$ while keeping $N_{s}$ fixed. We obtain
\begin{equation}
\begin{split}
D_{\text{gen}}&:=D\left(  \hat{\rho}_{RI}^{(0)}||\hat{\rho}_{RI}^{(1)}\right)
\\&=\frac{\kappa C^{2}}{N_b}\ln\left(  1+\frac{1}{N_{s}}\right)  +\mathcal{O}%
\left(  N_b^{-2}\right)  , \label{arbGaussrelent}%
\end{split}
\end{equation}%
\begin{equation}
\begin{split}
V_{\text{gen}}&:=V\left(  \hat{\rho}_{RI}^{(0)}||\hat{\rho}_{RI}^{(1)}\right)
\\&=\frac{\kappa C^{2}(2N_{s}+1)}{N_b}\ln^{2}\left(  1+\frac{1}{N_{s}}\right)
+\mathcal{O}\left(  N_b^{-2}\right)  . \label{arbGaussrelentvar}%
\end{split}
\end{equation}
For coherent states these quantities take the form
\begin{equation}
D_{\text{CS}}:=D\left(  \hat{\rho}_{\text{CS}}^{(0)}||\hat{\rho}_{\text{CS}%
}^{(1)}\right)  =\kappa N_{s}\ln\left(  1+\frac{1}{N_b}\right)  ,
\end{equation}%
\begin{equation}
V_{\text{CS}}:=V\left(  \hat{\rho}_{\text{CS}}^{(0)}||\hat{\rho}_{\text{CS}%
}^{(1)}\right)  =\kappa N_{s}(2N_b+1)\ln^{2}\left(  1+\frac{1}{N_b%
}\right)  ,
\end{equation}
which hold for all values of $N_{s}$, $N_b$ and $\kappa$. For comparative purposes we evaluate again to first order
in $N_b$ while keeping $N_{s}$ fixed to obtain the simple expressions
\begin{equation}
D_{\text{CS}}\simeq\gamma+\mathcal{O}\left(  N_b^{-2}\right)  ,~V_{\text{CS}%
}\simeq2\gamma+\mathcal{O}\left(  N_b^{-2}\right)  ,
\label{coherentrelentNb}%
\end{equation}
where%
\begin{equation}
\gamma:=\frac{\kappa N_{s}}{N_b}%
\end{equation}
is the signal-to-noise ratio (SNR), usually expressed in decibels (dB) via
$\gamma_{\mathrm{dB}}=10\log_{10}\gamma$.

In the limit of very large $M$, we can approximately neglect the variance
contribution and just consider the relative entropy in the type-II (missed
detection) error probability of Eq.~(\ref{typeIIerrorexpform}) which simply
becomes $P_{\text{md}}\simeq\exp\left[  -MD\left(  \hat{\rho}_{0},\hat{\rho
}_{1}\right)  \right]  $. Then, we can deduce that, for large $M$ and a very
high background $N_b\gg1$, the error exponent of a quantum-correlated source
[Eq.~(\ref{arbGaussrelent})] has the following ratio with respect to a
coherent state source [Eq.~(\ref{coherentrelentNb})]
\begin{equation}
A(C,N_{s}):=\frac{D_{\text{gen}}}{D_{\text{CS}}}=\frac{C^{2}}{N_{s}}\ln\left(
1+\frac{1}{N_{s}}\right)  .\label{asymmetricadvantagefunction}%
\end{equation}
\begin{figure}[h]
\vspace{+0.1cm} \centering
\includegraphics[width=8.6cm]{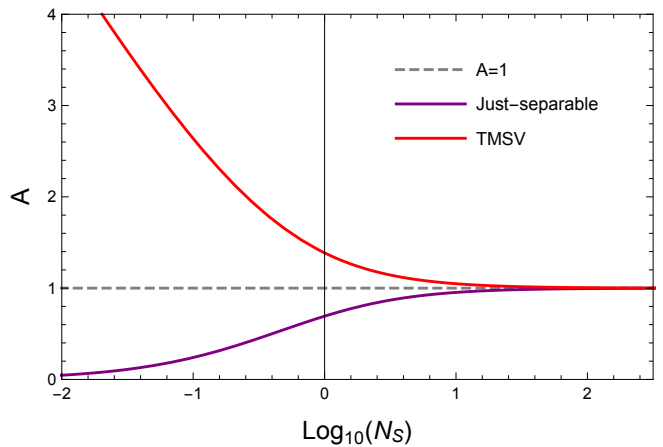} \caption{Error exponent ratio $A$ of
Eq.~(\ref{asymmetricadvantagefunction}) is shown as a function of number of
signal photons per mode, $N_{s}$. The just-separable discordant source quickly
approaches the coherent-state transmitter ($A=1$) and the TMSV source, already
at $N_{s}\simeq20$ photons. For increasing $N_{s}$, the ratio $A$
asymptotically approaches $1$ independent of the source specification.}%
\label{AsymmetricAdvantageFunctionALL}%
\end{figure}

In Fig.~\ref{AsymmetricAdvantageFunctionALL} we plot the ratio $A$ for a
just-separable discordant source ($C=N_{s}$) and that for a TMSV state, for
varying $N_{s}$. We can see how the ultimate benefits of employing maximal
entanglement for QI are exhibited only for very small energies, i.e., when
$N_{s}$ is of the order of units or less. For increasing $N_{s}$, the ratio
$A$ tends to the same asymptotic value, irrespective of source specification.
This also means that the just-separable source quickly approaches the
performance of QI at as little as about 20 photons per mode. As we will see in Sec.~\ref{SECroc}, for a given range $R$ the radar equation imposes a $1/R^4$ loss factor in received signal power and necessitates high overall photon numbers, particularly at long range, regardless of the underlying detection protocol. Our results show that at long range there is little-to-no advantage in using a QI-based radar over a coherent state protocol due to the need of large signal power ($N_{s}$). QI is thus limited to applications where losses are relatively small so $N_{s}$ may in turn take small values, for example, at short ranges.

\subsection{Receiver operating characteristic}\label{SECroc}

In the asymmetric setting, we now study the mis-detection probability versus
the false alarm probability of the generic Gaussian source with respect to the
classical benchmark of coherent states. For the latter, we consider the
performance achievable\ by coherent states and homodyne detection at the
output. This is the best-known measurement design which can be used when the
phase of the optical field is perfectly maintained in the interaction with the
target, so that one can adopt a coherent integration of the pulses (i.e., the
quadrature outcomes can be added before making a classical binary test on the
total value). If the phase of the field is deterministically changed to some
unknown value but it is still coherently maintained among the pulses, then the
typical choice is the heterodyne detection, followed by coherent
integration of the outcomes from both the quadratures. If the coherence is
lost among the pulses, then the classical strategy is to use heterodyne and
perform a non-coherent integration of the pulses, which means to sum the
recorded intensities (squared values of the quadratures). In this case, the
performance (for non-fluctuating targets) is given by the Marcum's
Q-function~\cite{Marcum}, an approximation of which is known as Albersheim's
equation~\cite{albersheim1981closed,richards2005fundamentals}. An
overestimation of the Marcum benchmark can be simply achieved by assuming a
single coherent pulse with mean number of photons equal to $MN_{s}$.

In mathematical terms, the ROC $P_{\text{md}}=P_{\text{md}}(P_{\text{fa}}%
)$\ of the generic Gaussian source can be upper bounded $P_{\text{md}}%
\leq\tilde{P}_{\text{md}}$ by combining Eqs.~(\ref{ROCub}),
(\ref{arbGaussrelent}) and~(\ref{arbGaussrelentvar}). For sufficiently large
$M$ (e.g., $\gtrsim10^{7}$), the second-order asymptotics is a good
approximation, and for large $N_b$ (e.g., $\gtrsim10^{2}$) the expansions in
Eqs.~(\ref{arbGaussrelent}) and~(\ref{arbGaussrelentvar}) are valid.
Therefore, under these assumptions, we may write
\begin{align}
\tilde{P}_{\text{md}}^{\text{gen}} &  =\exp\left\{  -\left[  \sqrt
{\frac{M\gamma}{N_{s}}}\Lambda C\ln\left(  1+\frac{1}{N_{s}}\right)
+\mathcal{O}(N_b^{-1},1)\right]  \right\}  ,\\
\Lambda &  :=\left(  \sqrt{\frac{M\gamma}{N_{s}}}C+\sqrt{2N_{s}+1}\Phi
^{-1}(P_{\text{fa}})\right)  .
\end{align}

In the case of coherent states and homodyne detection (followed by coherent
integration and binary testing), the ROC\ is given by combining the following
expressions
\begin{align}
P_{\text{fa}}^{\text{hom}}(x) &  =\frac{1}{2}\operatorname{erfc}%
\left[  \frac{x}{\sqrt{M(2N_b+1)}}\right] ,\label{eqq1}\\
P_{\text{md}}^{\text{hom}}(x) &  =\frac{1}{2}\operatorname{erfc}\left[ \frac{M \sqrt{2 \kappa N_{s}} - x}{\sqrt{M(2N_b+1)}}\right]    ,\label{eqq2}%
\end{align}
where $\operatorname{erfc}(z):=1-2\pi^{-1/2}\int_{0}^{z}\exp(-t^{2})dt$ is the
complementary error function. Therefore we can invert Eq.~(\ref{eqq1}) and replace in Eq.~(\ref{eqq2}) to
derive the corresponding ROC.

Finally, as already mentioned, we can also write a lower bound to Marcum's
classical radar performance by assuming a single coherent state with mean
number of photons $MN_{s}$\ so that the total SNR is given by $M\gamma$. This
can be expressed as follows
\begin{equation}
P_{\text{md}}^{\text{Marcum}}=1-Q\left(  \sqrt{2M\gamma},\sqrt{-2\ln
P_{\text{fa}}}\right)  ,
\end{equation}
where the Marcum Q-function is defined as%
\begin{equation}
Q(x,y):=\int_{y}^{\infty}dt~te^{-(t^{2}+x^{2})/2}I_{0}(tx),
\end{equation}
with $I_{0}(.)$ being the modified Bessel function of the first kind of zero
order~\cite{Marcum}.

Before comparing the ROCs of the various transmitters, let us choose a
suitable regime of parameters for potential short-range applications (of the
order of $1$m, e.g., for security or biomedical applications), where $N_{s}$ need not be too large and a quantum advantage may be observed. By fixing some
specific radar frequency $\nu$ and the temperature $T$ of the environment, we
automatically fix the mean number of photons $N_b$ of the thermal
background. Thus, for $\nu=1$GHz (L band) and $T=290\mathrm{K}$ (room
temperature), we get $N_b\simeq6\times10^{3}$ photons (bright noise). Assume
broadband pulses, with $10\%$\ bandwidth ($100$MHz), so that their individual
duration is about $10$ns. If we use $M=10^{8}$ pulses then we have an
integration time of the order of $1$s, which is acceptable for slowly-moving
or still objects. Since we are interested in low-energy applications, assume
$N_{s}=1$ mean photon per pulse. What is left is an estimation of the SNR
$\gamma$ which comes from the overall transmissivity/reflectivity $\kappa$.

This remaining quantity can be estimated using the radar equation. This
equation expresses the power $P_{R}$ of the return signal in terms of the
signal power $P_{T}$ at the transmitter, the cross section $\sigma$ of the
target, the range $R$ of the target, and other parameters, such as the
transmit antenna gain $G$, the receive antenna collecting area $A_{R}$ and the form
factor $F$ which describes the transmissivity of the space between the radar
and the target. It takes the form~\cite{radarBOOK}
\begin{equation}
P_{R}=\frac{GF^{4}A_{R}\sigma}{(4\pi)^{2}R^{4}}P_{T}.\label{RadarEquation}%
\end{equation}
Here the factor $(4\pi)^{-2}R^{-4}$ accounts for the loss due to the pulse
propagating as a spherical wave (back and forth). This is partly mitigated by
the gain $G$ which introduces anisotropies from the spherical wave
description, accounting for the directivity of the actual outgoing beam. In
fact $G$ describes the ratio between the power irradiated in the direction of
the target over the power that would have been irradiated by an isotropic
antenna~\cite{radarBOOK}. For a pencil beam, $G$ can be much higher than $1$
(which is the value of an isotropic antenna).

\begin{figure}[t]
\vspace{0.2cm}
\centering
\includegraphics[width=8.6cm]{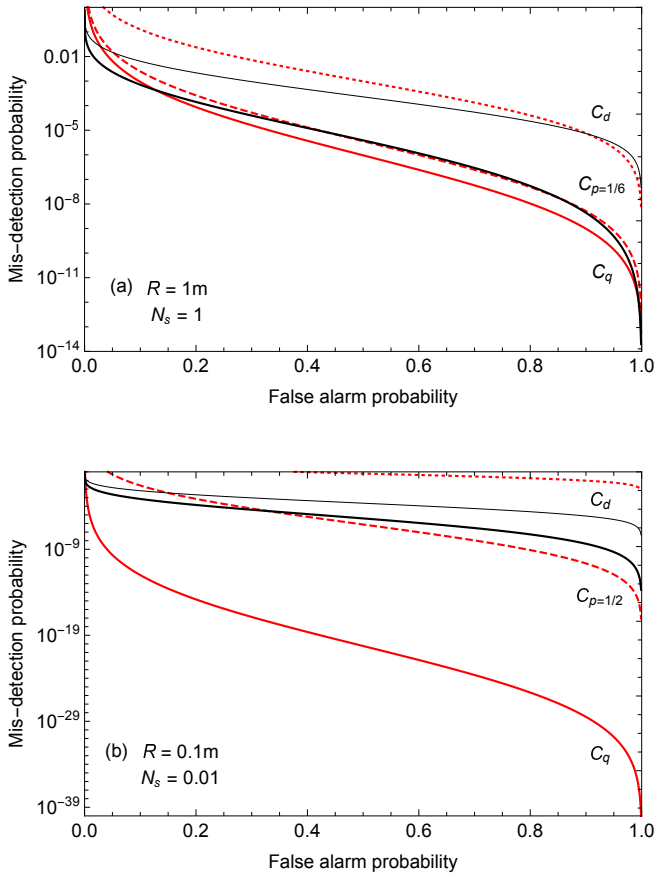}
\caption{Receiver operating
characteristics (ROCs) of the various setups. We show the upper bound
$\tilde{P}_{\text{md}}=\tilde{P}_{\text{md}}(P_{\text{fa}})$ for quantum illumination based on a
generic Gaussian state with off-diagonal correlation parameter $C(p)$ ranging
from the maximally-entangled state (red solid) to the just separable state
(red dotted). Between these two extremal curves, there are all the Gaussian
states with intermediate correlations. In particular, we show the performance
for an intermediate value of $p$ (red dashed). For comparison, we plot the ROC $P_{\text{md}}=P_{\text{md}%
}(P_{\text{fa}})$\ of the classical benchmark of coherent states plus homodyne
detection (black thick) and the lower bound to Marcum's classical performance
(black thin). Parameters are: $\nu=1$GHz and $T=290\mathrm{K}$\ (so that
$N_b\simeq6\times10^{3}$), $M=10^{8}$ pulses and
$\gamma_{\mathrm{dB}}=-70$dB. Upper panel (a): $N_{s} = 1$, corresponding to a range $R\simeq1$m with intermediate $p=1/6$; lower panel (b): $N_{s}=0.01$ corresponding to a range $R \simeq 0.1$m with intermediate $p=1/2$ and an integration time of about $1$s at $10\%$\ bandwidth ($100$MHz) in both cases.}%
\label{ROCpic}%
\end{figure}

It is clear that $\kappa$ also provides the ratio between received and
transmitted power, so that Eq.~(\ref{RadarEquation}) leads to%
\begin{equation}
\kappa=\frac{P_{R}}{P_{T}}=\frac{GF^{4}A_{R}\sigma}{(4\pi)^{2}R^{4}},
\end{equation}
which is also easy to invert, so as to express the range $R$ in terms of
$\kappa$ and the other parameters. Assume $F=1$ (no free-space loss) and an
ideal pencil beam, such that its solid angle $\delta$ is exactly subtended by
the target's cross section $\sigma$ (valid assumption at short ranges). This
means that gain is ideally given by%
\begin{equation}
G=\frac{4\pi}{\delta}=\frac{4\pi R^{2}}{\sigma},
\end{equation}
which fully compensates the loss in the forward propagation. Therefore, we
find%
\begin{equation}
\kappa=\frac{A_{R}}{(4\pi R)^{2}},~~R=\frac{1}{4\pi}\sqrt{\frac{A_{R}}{\kappa
}}.\label{ideall}%
\end{equation}
By fixing the receive antenna collecting area $A_{R}$, we have a one-to-one
correspondence between range $R$ and transmissivity $\kappa$. Assuming
$A_{R}=0.1$m$^{2}$ and short-range $R\simeq1$m, we get $\kappa\simeq
6\times10^{-4}$ which leads to $\gamma_{\mathrm{dB}}=-70$dB when we account
for the values of $N_{s}$ and $N_b$.

Considering this regime of parameters, we find the ROCs plotted in
Fig.~\ref{ROCpic}.\ In particular, we show the performance of a generic Gaussian source
with correlation parameter $C(p)=pC_{d}+(1-p)C_{q}$ between the extremal
points given by the just-separable source $C_{d}=N_{s}$\ and the
maximally-entangled source (at that energy) $C_{q}=\sqrt{N_{s}(N_{s}+1)}$
(another study of the ROC of the maximally-entangled case can be found in
Ref.~\cite{QuntaoOSA} but for the regime $N_{s}\ll1$). We perform the comparison for two scenarios while maintaining the same SNR, background characteristics and total number of uses: the first with $N_{s}=1$ and the second with $N_{s}=0.01$ corresponding to ranges $R=1$m and $R=0.1$m, respectively. From the figure, we can see that intermediate values of entanglement are able to beat the classical
benchmark given by coherent states and homodyne detection. The potential advantage is greater at lower signal energy $N_{s}$ or, equivalently, shorter range $R$ and additionally, the intermediate level of entanglement required in order to attain such an advantage reduces. In the upper panel, we consider out suggested upper limit for both range and signal energy: $R=1$m and $N_{s} =1$. This plot shows that though maximal entanglement is not strictly necessary for a quantum advantage, the scope for such an advantage is limited with the minimum intermediate level at $p=1/6$, i.e., very close to the maximally-entangled case. The lower panel highlights the benefits afforded to QI by limiting applications to short range and low signal energy, plotting results for $R=0.1$m and $N_{s}=0.01$. Here the minimum intermediate level is given by $p=1/2$ which yields a large range of source specifications capable of achieving a quantum advantage with potential performances several orders of magnitude greater than the optimal classical protocol. This effect becomes greater still at progressively shorter ranges and lower energies. It is precisely in these cases where we find that QI is most suited and, in the likely scenario that there are inefficiencies associated with source generation, enhanced detection performance is still achievable to a potentially very high degree.

\section{Conclusion}

In this work, we have investigated how to loosen the transmitter requirements
of QI, from the usual maximally-entangled TMSV\ source to a more general
quantum-correlated Gaussian source, which may become just-separable. At the
same time, we maintain the optimal quantum joint-measurement procedure at the
receiver side. We perform this investigation in both scenarios of symmetric
and asymmetric testing where we test the quantum performance with respect to
suitable classical benchmarks. Our results show that we can still find quantum
advantage by using Gaussian sources which are not necessarily maximally
entangled. In particular, this is an advantage which appears at short ranges,
so that the spherical beam spreading does not involve too many dBs of loss, a
major killing factor for any quantum radar design based on the exploitation of
quantum correlations.

A short-range low-power radar is potentially interesting not only
as a non-invasive scanning tool for biomedical applications but also for security
and safety purposes, e.g., as a scanner for metallic objects or as proximity
sensor for obstacle detection. Once
quantum advantage in detection is achieved at fixed target distance, it can be extended to variable distances to enable a measurement of the range. For instance, this could be done by sending
signal-idler pulses at different carrier frequencies and interrogating their reflection at different
round-trip times. For slowly moving objects at short ranges, the total interrogation time would
be small and the effective distance of the object could be well-resolved by sweeping a reasonable number of frequencies. In a static setting, e.g., biomedical, detection is naturally associated
with a fixed depth, which is then gradually increased so as to provide a progressive scan of the
target region. This quantum scanner would investigate the presence of the target at different
layers, e.g., of a tissue, while irradiating small energies. In conjunction with standard Doppler
techniques it could also extract information about the local velocity of the target within a layer
of the tissue. Such a quantum scanner would then realize an ideal non-invasive diagnostic tool.

\section*{Acknowledgements}
A.K. acknowledges sponsorship by EPSRC Award No. 1949572 and Leonardo UK. G.S. acknowledges sponsorship by European Union's Horizon 2020 Research and Innovation Action under grant agreement No. 745727 (Marie Sklodowska-Curie Global Fellowship `quantum sensing
for biology', QSB). Q.Z. is supported by the Army Research Office under Grant Number W911NF-19-1-0418 and Office of Naval Research under Grant Number N00014-19-1-2189. S.P. acknowledges sponsorship by European Union's Horizon 2020 Research and Innovation Action under grant agreement No. 862644 (`Quantum readout techniques and technologies', QUARTET).


\appendix

\section{Quantum Bhattacharyya bound for generic Gaussian source\label{App1}}

Using the generic quantum-correlated Gaussian source defined in Sec.~\ref{SECgen} along with formulae and tools for symmetric QHT described in Sec.~\ref{SECsymdet} of the main text, one can compute the exact form of the quantum Bhattacharyya bound (QBB) for QI. The complete formula is too long to be displayed however, imposing parameter constraints, one can achieve a closed-form for the asymptotic performance in specified limits.

We begin by assuming the typical conditions of QI, which are low-reflectivity
$\kappa\ll1$, high thermal-noise $N_b\gg1$, low photon number per mode
$N_{s}\ll1$. Then, using numerical techniques, one can confirm that using a TMSV state, the minimum error
probability satisfies~\cite{tan2008quantum}
\begin{equation}
P_{\text{err}}^{\text{TMSV}}\leq e^{-M\kappa N_{s}/N_b}%
/2,\label{bhattaqibound2}%
\end{equation}
which is exponentially tight in the limit of large $M$ and is valid under the parameter constraints previously defined.

\begin{figure}[t]
\vspace{+0.2cm}
    \centering
    \includegraphics[width=6cm]{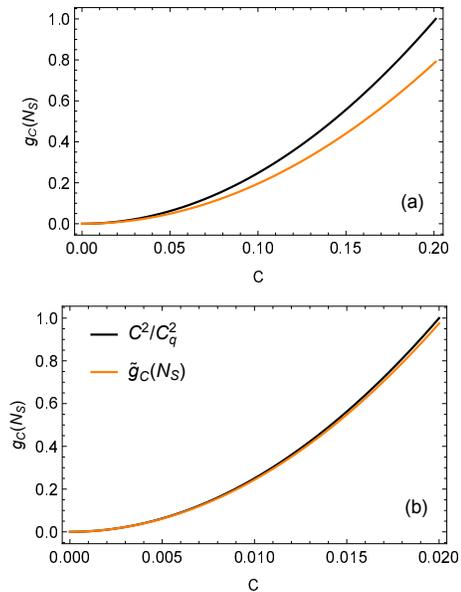}
    \caption{Numerical study of generic Gaussian source's QI error exponent for parameter values: (a) $N_{s}=10^{-2}$, $N_b=20$ and (b) $N_{s}=10^{-4}$, $N_b=200$. The plots confirm that for small $N_{s}$ and large $N_b$ we have that $\tilde{g}_C(N_{s}) \rightarrow g_C(N_{s}) = C^2 /C_{q}^{2}$ and our formula for generic Gaussian QI holds in this regime.}
    \label{fig:gensourcefactor}
\end{figure}

In order to extend Eq.~(\ref{bhattaqibound2}) to the error probability for a
generic Gaussian source, we first note that as we only vary the value of cross-correlation parameter $C$ the variation in the bound will be entirely dependent on this parameter. We also note that the parameter $C$ is constrained by the terms on the leading diagonal such that
\begin{equation}
0 \leq C \leq \sqrt{N_{s}(N_{s}+1)} := C_q,
\end{equation}
where the upper bound corresponds to the maximally-entangled TMSV state yielding Eq.~(\ref{bhattaqibound2}). Since the just-separable state corresponds to $C=C_d := N_{s}$ it is clear that $C=C(N_{s})$.

The form of Eq.~(\ref{bhattaqibound2}) is not surprising; the error exponent is directly proportional to the SNR, $\gamma=\kappa N_{s}/N_b$, and one would expect the same for our generic source. Starting with single probing, $M=1$, and subject to the limits $\kappa \ll 1$, $N_b\gg 1$ and $N_{s}\ll1$, we can write the QBB for our generic source as
\begin{equation}
P_{\text{err}}^{\text{gen}}\leq e^{-\kappa N_{s}g_{C}(N_{s})/N_b}/2,
\label{regenbound}
\end{equation}
where we define the function $g_{C}(N_{s})$ as a constant of proportionality, entirely dependent on the parameter $C$ and thus $N_{s}$. In particular, we demand the equivalence of exponents in the TMSV limit $C\rightarrow C_q$ such that $g_{C_q}(N_{s}) =1$, recovering the bound given by Eq.~(\ref{bhattaqibound2}).

To determine the form of $g_{C}(N_{s})$ we use a numerical program to perform an asymptotic expansion of our generic source's exact QBB for small $\kappa \ll 1$. Keeping terms to first order, we obtain an equation of the form
\begin{equation}
2 P_{\text{err}}^{\text{gen}} \leq 1-x\kappa + \mathcal{O}(\kappa^2) \simeq e^{-x\kappa},
\end{equation}
where the last equality holds when $x=\frac{N_{s}}{N_b} g_{C}(N_{s})$, from Eq.~(\ref{regenbound}), is small, i.e., $N_{s}\ll1$ and $N_b \gg1$.

Numerical analysis shows that the coefficient $x$ is exactly proportional to $C^2$, independent of $N_{s}$ and $N_b$, thus we can write $g_{C}(N_{s})\propto C^2$ and, imposing the condition that $g_{C_q}(N_{s}) =1$ determine that
\begin{equation}
    g_{C}(N_{s}) = C^2/C_{q}^{2}.
    \label{gCform}
\end{equation}

Fig. \ref{fig:gensourcefactor} plots the function $g_C(N_{s})$ as a function of cross-correlation parameter $C$ for two sets of parameter values: (a) $N_{s}=10^{-2}$, $N_b=20$ and (b) $N_{s}=10^{-4}$, $N_b=200$. It shows that in the regime of low brightness and high background the function $\tilde{g}_{C}(N_{s}) \rightarrow g_{C}(N_{s})$, given by Eq.~(\ref{gCform}), and we can write that the QBB for a generic Gaussian source is given by
\begin{equation}
P_{\text{err}}^{\text{gen}}\leq e^{-M\kappa N_{s}C^{2}/N_bC_{q}^{2}}/2,
\label{gensourcebound2}%
\end{equation}
as given in the main text. Note that the extension from $M=1$ to generic $M$ just follows from the structure
of the QCB and QBB in Eqs.~(\ref{QCBtext}) and~(\ref{QBBtext}).


\begin{thebibliography}{99}                                                                                               %


\bibitem {lehmann2006testing}E. L. Lehmann, and J. P. Romano, \textit{Testing
statistical hypotheses} (Springer Science \& Business Media, 2006).

\bibitem {Helstrom}C. W. Helstrom, \textit{Quantum Detection and Estimation
Theory}, \textit{Mathematics in Science and Engineering}, Vol. 123 (Academic
Press, New York, 1976).

\bibitem {Cover}T. M. Cover and J. A. Thomas, \textit{Elements of Information
Theory} (2n edition, Wiley, 2006).

\bibitem {watrous2018quantum}J. Watrous, \textit{The theory of quantum
information} (Cambridge University Press, Cambridge, 2018).

\bibitem {richards2005fundamentals}M. A. Richards, \textit{Fundamentals of
radar signal processing} (Tata McGraw-Hill Education, 2005).

\bibitem {lloyd2008enhanced}S. Lloyd, \textit{Enhanced sensitivity of
photodetection via quantum illumination}, Science \textbf{321}, 1463-1465 (2008).

\bibitem {tan2008quantum}S.-H. Tan\textit{ et al.}, \textit{Quantum
illumination with Gaussian states}, Phys. Rev. Lett. \textbf{101}, 253601 (2008).

\bibitem {barzanjeh2015microwave}S. Barzanjeh \textit{et al.},
\textit{Microwave quantum illumination}, Phys. Rev. Lett. \textbf{114}, 080503 (2015).

\bibitem {reviewSENSING}S. Pirandola, B. Roy Bardhan, T. Gehring, C.
Weedbrook, and S. Lloyd, \textit{Advances in Photonic Quantum Sensing}, Nat.
Photon. \textbf{12}, 724-733 (2018).

\bibitem {FFSFG}Q. Zhuang, Z. Zhang, and J. H. Shapiro, \textit{Optimum
mixed-state discrimination for noisy entanglement-enhanced sensing}, Phys.
Rev. Lett. \textbf{118}, 040801 (2017).

\bibitem{lopaevaexp}E. D. Lopaeva, I. Ruo Berchera, I. P. Degiovanni, S. Olivares, Giorgio Brida, and Marco Genovese, \textit{Experimental realization of quantum illumination}, Phys. Rev. Lett. \textbf{110}, 153603 (2013).

\bibitem{agudelo}E. Agudelo, J. Sperling, and W. Vogel, \textit{Quasiprobabilities for multipartite quantum correlations of light}, Phys. Rev. A \textbf{87}, 033811 (2013).

\bibitem{sperling}J. Sperling, M. Bohmann, W. Vogel, G. Harder, B. Brecht, V. Ansari, and C. Silberhorn, \textit{Uncovering quantum correlations with time-multiplexed click detection}, Phys. Rev. Lett. \textbf{115}, 023601 (2015).

\bibitem{shahandeh}F. Shahandeh, A. P. Lund, and T. C. Ralph, \textit{Quantum correlations in nonlocal boson samplinh}, Phys. Rev. Lett. \textbf{119}, 120502 (2017).

\bibitem{spedalieri}G. Spedalieri, C. Lupo, S. L. Braunstein, and S. Pirandola, \textit{Thermal quantum metrology in memoryless and correlated environments}, Quantum Science and Technology \textbf{4}, 015008 (2018).

\bibitem {RMP}C. Weedbrook, S. Pirandola, R. Garcia-Patron, N. J. Cerf, T. C.
Ralph, J. H. Shapiro, and S. Lloyd, \textit{Gaussian Quantum Information},
Rev. Mod. Phys. \textbf{84}, 621 (2012).

\bibitem {EntBreak}S. Pirandola, \textit{Entanglement Reactivation in
Separable Environments}, New J. Phys. \textbf{15}, 113046 (2013).

\bibitem{ModiDiscord}K. Modi, A. Brodutch, H. Cable, T. Paterek and V. Vedral, \textit{The classical-quantum boundary for correlations: Discord and related measures}, Rev. Mod. Phys. \textbf{84}, 1655-1707 (2012).

\bibitem {OptDisc}S. Pirandola, G. Spedalieri, S. L. Braunstein, N. J. Cerf,
and S. Lloyd, \textit{Optimality Gaussian discord}, Phys. Rev. Lett.
\textbf{113}, 140405 (2014).

\bibitem {Gdis1}P. Giorda and M. G. A. Paris, \textit{Gaussian Quantum
Discord}, Phys. Rev. Lett. \textbf{105}, 020503 (2010).

\bibitem {Gdis2}G. Adesso and A. Datta, \textit{Quantum versus Classical
Correlations in Gaussian States}, Phys. Rev. Lett. \textbf{105}, 030501 (2010).

\bibitem {PLOB}S. Pirandola, R. Laurenza, C. Ottaviani, and L. Banchi,\textit{
Fundamental Limits of Repeaterless Quantum Communications}, Nat. Commun.
\textbf{8}, 15043 (2017). See also arXiv:1510.08863 (2015).

\bibitem {helstrom1969quantum}C. W. Helstrom, \textit{Quantum detection and
estimation theory}, J. of Stat. Phys \textbf{1}, 231-252 (1969).

\bibitem {QCB}K. M. R. Audenaert, J. Calsamiglia, L. Masanes, R. Munoz-Tapia,
A. Acin, E. Bagan, and F. Verstraete, \textit{Discriminating States: The
Quantum Chernoff Bound}, Phys. Rev. Lett. \textbf{98}, 160501 (2007).

\bibitem {pirandola2008computable}S. Pirandola and S. Lloyd,
\textit{Computable bounds for the discrimination of Gaussian states}, Phys.
Rev. A \textbf{78}, 012331 (2008).

\bibitem {serafini2003symplectic}A. Serafini, F. Illuminati, and S. De Siena,
\textit{Symplectic invariants, entropic measures and correlations of Gaussian
states}, J. of Phys. B \textbf{37}, L21 (2003).

\bibitem {pirandola2009correlation}S. Pirandola, A. Serafini, and S. Lloyd,
\textit{Correlation matrices of two-mode bosonic systems}, Phys. Rev. A
\textbf{79}, 052327 (2009).

\bibitem {hiai1991proper}F. Hiai and D. Petz, \textit{The proper formula for
relative entropy and its asymptotics in quantum probability}, Commun. Math.
Phys. \textbf{143}, 99-114 (1991).

\bibitem {ogawa2005strong}T. Ogawa and H. Nagaoka, \textit{Strong converse and
Stein's lemma in quantum hypothesis testing}, Asymptotic Theory Of Quantum
Statistical Inference: Selected Papers, 28-42 (World Scientific, 2005).

\bibitem {QHBound}K. M. R. Audenaert, M. Nussbaum, A. Szkola, and F.
Verstraete, \textit{Asymptotic Error Rates in Quantum Hypothesis Testing},
Commun. Math. Phys. \textbf{279}, 251 (2008).

\bibitem {GaeGAUSS}G. Spedalieri and S. L. Braunstein, \textit{Asymmetric
quantum hypothesis testing with Gaussian states}, Phys. Rev. A \textbf{90},
052307 (2014).

\bibitem {li2014second}K. Li, \textit{Second-order asymptotics for quantum
hypothesis testing}, Annals of Statistics \textbf{42}, 171-189 (2014).

\bibitem {BanchiPRL}L. Banchi, S. L. Braunstein, and S. Pirandola,
\textit{Quantum fidelity for arbitrary Gaussian states}, Phys. Rev. Lett.
\textbf{115}, 260501 (2015).

\bibitem {RevQKD}S. Pirandola, U. L. Andersen, L. Banchi, M. Berta, D.
Bunandar, R. Colbeck, D. Englund, T. Gehring, C. Lupo, C. Ottaviani, J.
Pereira, M. Razavi, J. S. Shaari, M. Tomamichel, V. C. Usenko, G. Vallone, P.
Villoresi, and P. Wallden, \textit{Advances in Quantum Cryptography},
arXiv:1906.01645 (2019).

\bibitem{LaurenzaBounds}R. Laurenza, S. Tserkis, L. Banchi, S.L. Braunstein, T.C. Ralph, and S. Pirandola, \textit{Tight bounds for private communication over bosonic Gaussian channels based on teleportation simulation with optimal finite resources}, Phys. Rev. A, \textbf{100}, 042301 (2019); M. M. Wilde, M. Tomamichel, S. Lloyd, and M. Berta, \textit{Gaussian Hypothesis Testing and Quantum Illumination}, Phys. Rev. Lett. \textbf{119}, 120501 (2017).


\bibitem {radarBOOK}I. S. Merrill \textit{et al.}, \textit{Introduction to
radar systems} (McGraw-Hill, 1981).

\bibitem {Marcum}J. I. Marcum, \textit{A Statistical Theory of Target
Detection by Pulsed Radar: Mathematical Appendix},\ RAND Corporation, Santa
Monica, CA, Research Memorandum RM-753, July 1, 1948. Reprinted in IRE
Transactions on Information Theory \textbf{IT-6}, 59--267 (1960).

\bibitem {albersheim1981closed}W. Albersheim, \textit{A closed-form
approximation to Robertson's detection characteristics}, Proc. of the IEEE
\textbf{69}, 839-839 (1981).

\bibitem {QuntaoOSA}Q. Zhuang, Z. Zhang, and J. H. Shapiro,
\textit{Entanglement-enhanced Neyman--Pearson target detection using quantum
illumination}, J. Opt. Soc Am. B \textbf{34}, 1567-1572 (2017).
\end{thebibliography}
\end{document}